# On-Demand Grid Provisioning Using Cloud Infrastructures and Related Virtualization Tools: A Survey and Taxonomy

[1*]Shafi'i Muhammad Abdulhamid, [2]Muhammad Shafie Abd Latiff and [3]Mohammed Bakri Bashir

[1,2,3]Faculty of Computing, Universiti of Teknologi Malaysia, 81310 Skudai, Jahor, Malaysia

[1]Department of Cyber Security Science, Federal University of Technology Minna, Nigeria

*Abstract* - **Recent researches have shown that grid resources can be accessed by client on-demand, with the help of virtualization technology in the Cloud. The virtual machines hosted by the hypervisors are being utilized to build the grid network within the cloud environment. The aim of this study is to survey some concepts used for the on-demand grid provisioning using Infrastructure as a Service Cloud and the taxonomy of its related components. This paper, discusses the different approaches for on-demand grid using infrastructural Cloud, the issues it tries to address and the implementation tools. The paper also, proposed an extended classification for the virtualization technology used and a new classification for the Grid-Cloud integration which was based on the architecture, communication flow and the user demand for the Grid resources. This survey, tools and taxonomies presented here will contribute as a guide in the design of future architectures for further researches.**

## I. INTRODUCTION

The cloud computing technology is fast developing towards the direction of becoming the de facto standard of internet computing, storage and accommodating infrastructures, platforms and software both in the industries and the academia. The huge scalability prospects offered by cloud infrastructures can be easily harnessed not only for services and applications hosting but also as an on-demand grid computing resource. Cloud infrastructure system can be used to provide on-demand grid resources with the support of virtual machines (VM) and hypervisors otherwise known as virtual machine monitors [1, 2]. The main goal of this type of architecture is to offer an arbitrary complex grid infrastructures on-demand. This is to be able to provide for the demand needed by any given service, and at the same time taking advantage of the pay-per-use paradigm and the capabilities of the cloud computing model. It can be used to present solutions to mechanisms that are possibly used to meet up with a certain quality of service (QoS) or satisfy peak demands this service may require. In other words, it can be used to address the requirement to attend fluctuating peak demands in grid computing [3].

A virtual machine (VM) can be used to maintain different processes or a complete system subject to the abstraction level where virtualization occurs. Most VMs are compatible to elastic hardware usage and software segregation, whereas some translate from one instruction set to the next. A system VM offers a complete environment in which an operating system and many processes, probably belonging to various clients, may reside mutually. Using the VMs, a single-host hardware platform may be compatible with multiple, isolated guest operating system environments concurrently [4]. The use of virtualization, along with an efficient VMM, forms an innovative virtualization level that separates the service workload from the resource management [5, 6]. In system VM architecture, the virtualising software is built on top of the current host operating system, giving room to a hosted VM.

The benefit of a hosted VM is that the client mounts it just like a normal application program. In additional, virtualising software can now depend on the resident operating system to run device drivers and some lower-level services instead of on the virtualisers [4]. Virtualization infrastructures like the Xen [6], Kernel-based Virtual Machine (KVM) [7], Open Nebula [8] and VMWare [9] had introduced another level of abstraction called Hypervisor. A hypervisor runs at the topmost layer of the physical hard-ware system, sharing jobs to separate execution environments known as the VMs, which execute on a separate virtualised operating system. Hypervisor controls the activities of these virtualised operating systems, which includes; booting, suspending or shutting down systems as at when due. A good number of the hypervisors are capable of replication and migration of VMs without necessarily ending the task of the operating system. The basic enabler for Cloud computing is the split between resource provisioning and operating systems introduced by the virtualization systems. This is also the enabler for the virtualization of the Infrastructure-as-a-Service (IaaS) cloud [10].

Foster, et al. [11] defined grid concept as *"a coordinated resource sharing and problem solving in dynamic, multi-institutional virtual organizations".* Grid Infrastructure has a traditional approach to submitting, managing and executing tasks. In the grid design, isolated execution is still a problem because it offers different settings, like the operating systems, varied middleware design and again disparities in file system outputs in minimal usage of accessible resources by the clients. Many clients distribute and execute their tasks in the end resource, which distributes the same operating system of the resource, occasionally this leads to optimization problems. Grid computing constitutes various modules such as; client and administrator interface, workload management, data management, security, resource management, and scheduler. The strength of the grid computing system is not just in the combined computational power, data storage and network bandwidth. The grid computational system requires the QoS of job execution. One of the objectives of the grid is to compute huge number of applications with the support of virtualized resources [12].

Foster, et al. [13] also defined cloud as "*a large-scale distributed computing paradigm that is driven by economies of scale, in which a pool of abstracted, virtualized, dynamically-scalable, managed computing power, storage, platforms, and services are delivered on demand to external customers over the Internet".* VMs' growing omnipresence has enabled clients to build personalized environments on top of physical infrastructure and has simplified and increase the development of business paradigms in cloud computing. Virtualization helps to increase the development of cloud computing services.

Cloud computing services can be categorized into three: software-as-a-service (SaaS), platform-as-a-service (PaaS), and infrastructure-as-a-service (IaaS). In this case, we are more concern about IaaS, which focus on providing computing resources or storage as a service to clients [14, 15]. The IaaS cloud provide the needed VMs to be used in developing the infrastructural grid to be provisioned on-demand through the pay as you use cloud model.

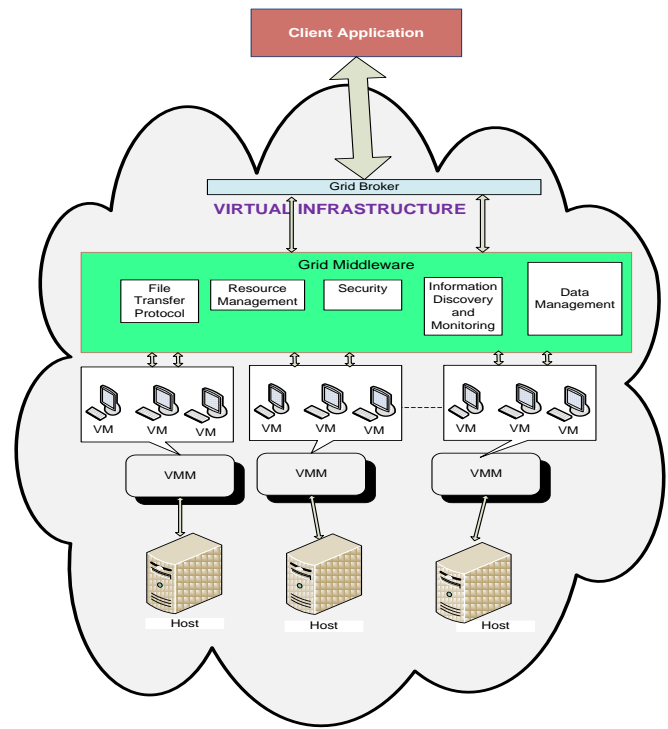

Fig. 1: A Virtual Infrastructural Cloud for On-Demand Grid Resources Provisioning

In the Fig. 1 above, it shows a virtual Infrastructure as a Service (IaaS) clouds that is used to provide a flexible and fast way to acquire VM as demand keeps fluctuating. The VM reside in the VMM, while VMM resides in the host operating system in the physical system. The grid middleware is installed in the VMs to able resources synchronization from heterogeneous administrative domains for job execution. In this case, clouds are used by grids as providers of devices like the middleware in the virtual machines, so that they only use the resources they need at particular time on demand. But this necessitates grids middleware to be able to decide when to allocate and release those resources [16, 17].

The objectives of this paper are to focus on the various concepts and tools used for on-demand grid resource provisioning using cloud infrastructure and the taxonomy of related components. Section II discusses the related works. Section III discusses the taxonomy of the virtualization tools used for the on-demand grid provisioning in cloud computing. The section IV presents a novel taxonomy for Grid-Cloud integration. While the section V details more discussion on the structure and functions of the various components used in the design of on-demand Grid provisioning using cloud infrastructure. The section VI discusses the possible economic impacts of the on-demand grid provisioning in cloud. The section VII is for discussion of results. Finally, section VIII concludes the paper with a summary of our contributions and future works.

## II. RELATED WORKS

Wang, et al. [18], proposed a novel method for Grid computing resource provisioning using VMs as computing resources and also use Virtual Distributed Environments (VDE) for Grid clients. They developed a Grid middleware called Grid Virtualization Engine to function as a building block for VMs in the Grids. The paper also introduced a use case on-demand to build a virtual e-Science infrastructure to validate the method used. However, the researchers only recommends application of this method as a future work on the IaaS cloud to provide on-demand grid resources, but never got to implement it that way. Varma and Choi [12] in their paper titled "*Extending Grid Infrastructure Using Cloud Computing*" used Globus toolkit middleware to extend the grid computing resources using eucalyptus cloud environment.

Virtualization in relation to Grid was realized by combining the GT4 and Eucalyptus features. The result shows that the extension of the Globus toolkit middleware with the Eucalyptus cloud environment will help the user to execute the jobs abstractedly with optimum deployment of the resources. However, the research does not address the issues related to efficient scheduling, resources management and fault tolerance.

Sahoo, et al. [19] provides a survey and taxonomy paper for virtualization in relation to reliability, availability and security issues. They described the fundamentals of virtualization technology and addressed the advantages and disadvantages of virtualization along with classification and challenges. Even though, the taxonomy was made based on system, storage and network, it fails to provide a class for virtualization in relation to cloud infrastructures. The classification does not also consider virtualization based on the on-demand grid provision using IaaS cloud. Caron, et al. [2] presented a proof of concept for using a Cloud infrastructure as computational resource through a Grid middleware. They demonstrated by using EUCALYPTUS (the open-source Cloud) as a resource for the DIET Grid middleware. Even though they underscored the most significant issues that arise when dealing with this subject, the research serves only as a proof of concept leaving open issues such as data management, scheduling and fault tolerance for future expansions.

The InterGrid system uses VMs as infrastructures to construct computational environments that cut across many locations. These types of environment can be extended to work on cloud infrastructures, such as EUCALYPTUS and Amazon EC2. Di Costanzo, et al. [14] presented an abstract view of the proposed architecture and its application. The research shows the scalability of an InterGrid-managed infrastructure and how it can take advantage of cloud infrastructures. However, they do not address security aspects in the work because they pointed out that these are to be tackled at the OS and network levels. Dornemann, et al. [20], in their paper titled "*On-Demand Resource Provisioning for BPEL Workflows Using Amazon's Elastic Compute Cloud*" presented an algorithm that automatically schedules workflow phases to idle hosts and offers new hosts using Cloud computing infrastructures in demanding periods.

The architecture does not need any alterations to the Business Process Execution Language (BPEL) standard. The experiment was carried out based on the ActiveBPEL engine and Amazon's EC2. However, a more complex algorithm is required to handle jobs with regard to data dependencies in between workflow phases, such that excessive data transfers is avoided.A cloud taxonomy was proposed by Abbadi [21] concentrating on infrastructure constituents communication, properties and management services. The paper also presented a case study of the critical application architecture by extending the proposed classification, and then derived the management services using the provided case study. However, this taxonomy does not include integrated cloud-grid infrastructures, properties or management.

Belgacem, et al. [22] presented two scenarios used to combined Grid/Cloud architecture for deploying phylogeny application known as MetaPIGA. The trick is to take advantage of Grid and Cloud infrastructures so as to build a high performance, reliable and open platform. The first scenario shows that, the Cloud infrastructures are not directly accessible by the clients; instead must pass through the Grid. While in the second case, clients submit tasks directly to the Cloud. These two scenarios have been verified using the MetaPIGA system combined with the Amazon, Azure and Venus-C Cloud infrastructures.

However, the research only recommended as a future work to generalize these scenarios to other applications and build a "generic" toolkit environment that will support other Cloud platforms. They did not consider a scenario where the grid will be built directly inside the cloud with the aid of virtualization technology.

### III. TAXONOMY OF VIRTUALIZATION TOOLS

Basically, there are two components of virtualization tools to be used when dealing with an on-demand grid provisioning. That is, the virtual machines or the guest operating systems and the hypervisors that manages and controls them on the physical host. This taxonomy tries to classify the hypervisors under the virtualization technology in relations to the on-demand grid resource provisioning using cloud computing.

Over the years, the virtualization technologies have undergone tremendous changes, especially after the introduction of cloud computing technology somewhere in 2008. Both public and private cloud computing operators begins to provide infrastructures as a service (IaaS) cloud through the provision of VMs with the help of hypervisors or VMMs. This development immediately changed the equation and call for the reclassification of the virtualization technologies. Here, we revisit and modify the taxonomy presented by Sahoo, et al. [19], with the aim of adapting it to suit the concept of grid resource provisioning using infrastructure as a service cloud. Fig. 2 below shows the adapted classification of the virtualization technology used for on-demand grid provisioning. It shows the Grid and Cloud virtualization classes added to the existing ones.

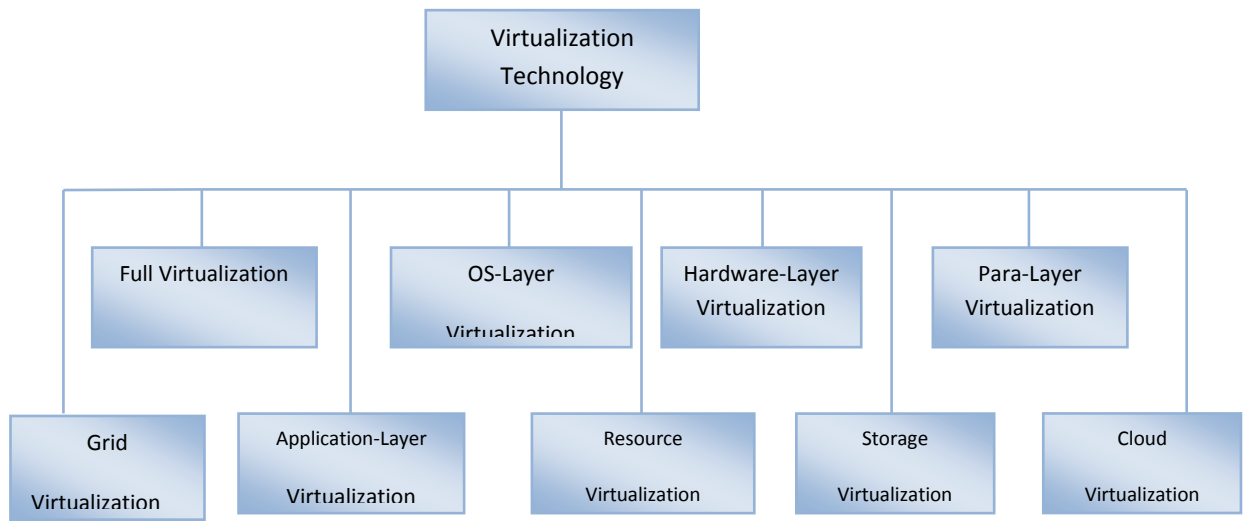

Fig. 2: Adapted Taxonomy of Virtualization Technology in On-Demand Grid Provisioning

#### A. Full Virtualization

Full virtualization is the only class that needs no hardware backing or operating system support to virtualize complex and privileged instructions. The VMM interprets all operating system directives on the fly and caches the outcomes for prospective usage, while user level directives executes unchanged at inherent speed. Some typical examples of full virtualization includes VMware's virtualization products and Microsoft Virtual Server [23].This category of virtualization technology is also known as VM on top manager and executes of a host OS, usually as an application in client space. The implication is that, in the virtual machines, the applications and the guest operating system executes on a virtual hardware via the hypervisor. Nevertheless, the VM environment that provides adequate illustrations of the primary hardware to permit guest OS to function without any adjustment can be considered to offer "*Full Virtualization*" [19].

#### B. OS-Layer Virtualization

The operating system-layer virtualization uses the guest OS in issuing codified hypercalls. This also runs on Xen and VMware. This is also called Single Kernel Image (SKI) or container-based virtualization, this approach executes virtualization technology by executing more instances of the same OS at the same time. This implies that not just the hardware but the host operating system is the one being projected. The subsequent VMs all use the same virtualized operating system image [24].

#### C. Hardware-Layer Virtualization

The hardware-layer virtualization exit to root mode on privileged instructions. It also has excellent compatibility with unmodified guest operating system. Some typical examples includes VMware, Microsoft, Parallels and Xen [23]. This category of virtualization is usually implemented on the server market because of its high VM segregation and performance. In this case, the hypervisor executes directly on hardware, managing and synchronizing the access of the guest operating systems to the hardware resources [25].

### D. Para-Layer Virtualization

Paravirtualization is main different between full virtualization and paravirtualization are that the unmodified operating system does not know it is virtualized and sensitive operating system calls are trapped using binary translation. A good example of para virtualization include the open source Xen project [6, 23]. In the case of Para-virtualization, the guest operating system that is executing should be adjusted in order to be operated in the virtualized settings. Para-virtualization is a subclass of server virtualization that uses a small application interface between the host hardware and the adapted guest operating system. The communication between devices in para virtualized setting is related to the device communication in a full virtualization. But the virtualized devices in para virtualized setting also depend on hardware device drivers of the principal host [19].

### E. Application-Layer Virtualization

In the Application-Layer virtualization, the client can be able to execute a server application locally with the aid of local resources without passing through the difficulty of installing this application on the machine. These types of virtualized applications are customized to function in a secluded virtual environment that contains only the resources required for the application to function. Therefore, each client of the application virtualization possesses remote virtualized application settings. The remote virtual environment functions as another level between the application and the host OS [26].

### F. Resource Virtualization

This is the act of virtualizing precisely system resources such as "storage volumes, name spaces and the network resources" is also referred to as resource virtualization. The following are some methods used to implement resource virtualization [27].

- Combining several discrete components into larger resource pool.

- High performance clusters of computers where many isolated systems are connected to form huge supercomputer with massive resources.
- Segregating a lone resource such as disk space into many smaller and easily available resources of same form.

### G. Storage Virtualization

Storage virtualization is another type of resource virtualization. In this case, a logical storage is generated by conceptualizing the entirely the physical storage resources that are distributed on the network. To start with, the physical storage resources are combined to produce storage pools that produce the logical storage. Then the logical storage is now combined with the distributed physical resources to form a single colossal storage device to the client [19].

### H. Grid Virtualization

Different grid middleware implemented virtualization of grid resources in different ways, but the common idea between them is the Virtual Organisation (VO) or the Virtual Organisation Clusters (VOC). The VOCs are an innovative way for overlaying dedicated private cluster systems on prevailing grid structures. They provide tailored, homogeneous working conditions on a per-VO basis, without the cost of physical cluster construction or the overhead of per-job containers [28]. Figueiredo [29] was the first to make a point for grid virtualization. The hardware administrators retain control of the VMM, coarse-grained resource sharing and monitoring can be executed on a per-VM basis, allowing hardware resources to be shared among different VMs.

### I. Cloud Virtualization

It is important to control layers, creating and monitoring the VMs, this requirement is been taken care of by Virtual Machine Monitor (VMM) or Hypervisor. This is one of the methods that makes possible to have many OSs ability in one single system [30, 31]. In recent times, the cloud computing paradigm influences virtualization technology and offers an on-demand resource provisioning over the World Wide Web on the basis of pay-per-use. This enables many organisations to cut-down their costs on maintenance of computing environment and subcontract the computational demands to the on-demand Cloud. As a result of this, virtualization now forms the foundation of Cloud Computing, as it provides the ability to combine computing resources from clusters or grids and vigorously provisioning of VM resources to clients on-demand [32, 33].

## IV. TAXONOMY OF GRID-CLOUD INTEGRATION

The popular integration processes of the Grid and Cloud computing are the "Grid on Cloud" and the "Cloud on Grid". The integration of Grid computing resources and virtual infrastructural cloud environment using VMs has been a very wide area of research and recently many approaches has been presented. But over the years, there have been little efforts to reclassify these different research approaches to the Grid-Cloud integration process to accommodate new architectures, user demand and communication between the Grid and the Cloud environments. Here we present a taxonomy that was based on the architectural approach, the communication and the user demand to the grid resources. This classification is divided into three; the disjointed, the partial and the full grid-cloud integration approach.

### A. Disjointed Grid-Cloud Integration

In this class of integration between the Grid environment and the Cloud infrastructures, the client makes use of an intermediate application to share the tasks between the grid and the cloud infrastructure. This is called *Disjointed Grid-Cloud integration* because; there is no direct link between the grid environment and the cloud infrastructures. The intermediate application can see directly both the grid and the cloud resources.

In the Fig. 3 below, it shows that grid user can only schedule jobs to the cloud indirectly through the intermediate application. It is the function of the application to format the jobs to fit in the traditional way of execution jobs in either grid or in the cloud as the case may. Researchers that have used this type of integration technique includes [22, 34]. This type of integration technique can also be used to share jobs between a remotely independent grid and another on-demand grid setup within a Cloud but using different types of middleware. In this case, when a job is been sent from the remotely independent grid, it first of all passes through the intermediate application. It is the function of the intermediate application to format the job so that it fits to the style of job execution in the other middleware. It is important to note here that the on-demand grid within the cloud cannot be able to communicate with the independent grid without the intermediate application [2, 3].

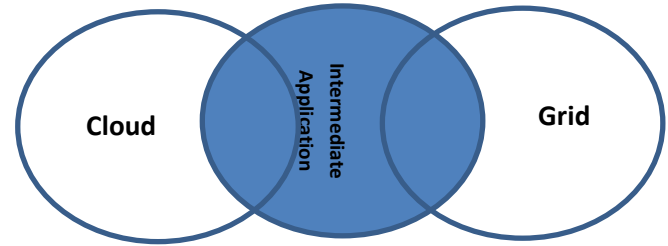

Fig.3: Disjointed Grid-Cloud Integration (Using an Intermediate Application)

### B. Partial Grid-Cloud Integration

In this category of Grid-Cloud integration, there exist a partial contact between the Grid environment and the Cloud infrastructures mostly through the extension of grid middleware. It is called *Partial Grid-Cloud integration* because; the grid user can be able to allocate jobs directly to the cloud at peak periods on-demand. It is sometime called extended or elastic on-demand grid. The user already has a partial grid for running his jobs and only need to use the cloud when dealing with delicate or critical jobs that cannot wait on the queue for too long without being executed (see Fig. 4 below).

The grid middleware is extended and configured in some of the VMs within the cloud infrastructure on-demand. This will give room for the traditional grid jobs format to be run within the cloud, while taking advantage of the pay-per-use nature of the cloud. The economic advantage is that, the grid user only pays for the short period he or she uses the cloud and therefore cost less compared to someone that run all his or her jobs in the cloud. Some of the researches that used this architectural approach to grid-cloud integration includes [12, 22, 35, 36].

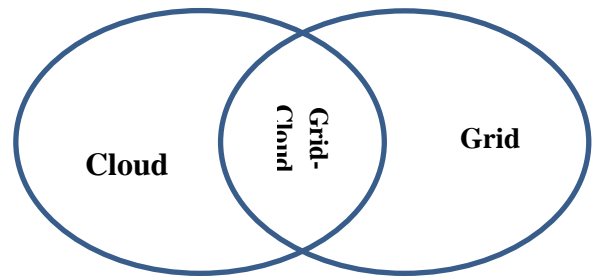

Fig. 4: Partial Grid-Cloud Integration (Extended On-Demand Grid)

### C. Full Grid-Cloud Integration

This class of integration technique is called *Full Grid-Cloud integration* because; the entire grid environment is built inside the cloud infrastructures. The cloud provides the VMs, the networking power and the pay-per-use capabilities, while the grid middleware is configured within the VMs with the support of the hypervisors. What makes a grid is the middleware, without it what we have is just a networking environment. It is important to note that here, the grid user do not have an existing grid to use. The user submits all his or her jobs to the on-demand grid within the cloud and also pays for all the execution time [2, 16] This architecture is also referred to as on-demand grid provisioning using IaaS cloud. This arrangement has the economic advantage of saving the users the cost of setting up a physical grid environment, as they may not need it. Therefore, the client will find it cheaper to pay for that particular job instead of setting up a physical grid. Fig.1and also Fig. 5 illustrates this concept. Some of the researches that have used this type of architecture includes [37-39].

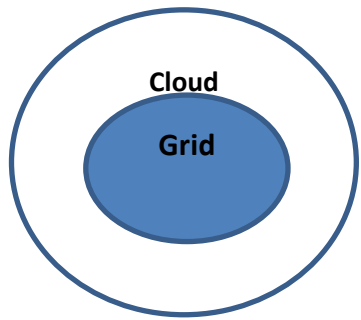

Fig. 5: Full Grid-Cloud Integration (On-Demand Grid Provisioning Using Cloud Infrastructure)

## V. ON-DEMAND GRID PROVISIONING USING CLOUD INFRASTRUCTURE

The term "On-demand" refers to the fact that client of the cloud resources only get access to them when they need the resources (in this case grid resources) , for how long they need them, and only pay for this actual usage [40]. One of the objectives of the on-demand grid provisioning is to use cloud powers in commercializing grid computing resources. The cloud infrastructures have the ability to provide VMs to clients as a service at a fee [41]. These VMs can now be used to setup a grid network within the cloud and issued out to customer on-demand. The Partial and Full Grid-Cloud integration classified above are good examples of on-demand grid resource provisioning using cloud infrastructures. The Fig. 6 below is a typical illustration of the architecture as demonstrated by [5, 28, 36, 37].

The Cloud computing infrastructures can either be a commercial cloud (like the Amazon EC2 [42]) or scientific cloud (like the Nimbus) or even open-source technologies (like the Eucalyptus and the Globus VWS). The cloud can as well be Internal which means private clouds (like the Eucalyptus) and External which means public clouds (like the Amazon EC2). Also, there are many Virtual Infrastructure Managements in IaaS cloud, these includes CLEVER, GoGrid, VOC, OpenNebula, and Globus Nimbus. These Cloud infrastructures are in turn used to accommodate the VMM, which in turn accommodate the VMs and the grid middleware [5, 43]. Virtualization is generally defined as a technology that presents a software abstraction layer between the hardware, the OS and applications running on top of it. The virtualization technologies often used to virtualize the cloud infrastructure and also manage the VMs (Guest OS) is the hypervisors. Some of the hypervisors used includes Xen [2], KVM [28], VMWare [44], OpenNebula [14] and VirtualBox. The hardware managers preserve exclusive control over the hypervisors, coarse-grained resource management can be employed on a per-VM basis, allowing hardware resources to be distributed among different VMs.

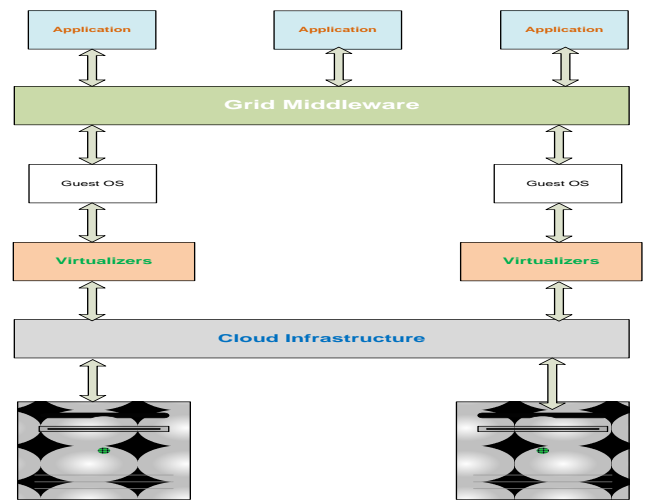

Fig. 6: On-Demand Grid Resource Provisioning in the Cloud Infrastructure

The Grid middleware which is a set of services and protocols that allows a collection of resources in grid. It provides a level of conception that hides variations in the primary technologies (e.g. computer clusters, storage managers, application services, etc.). The Grid middleware that is used within the VMs (Guest OS) is to provide traditional grid services to the on-demand grid clients. Some of the middleware services include information discovery and monitoring, resource management, security, Grid scheduling, load balancing, File Transfer Protocol (FTP) and data management. Some of the Grid middleware used in the previous works to include Globus [20, 44, 45], DIET-Solve [2], ASKALON [35], Condor [46], GEMBus [18],

BonjourGrid [47], g-Eclipse [48], UNICORE [49], Legion [49] and GridSAM [18]. Clearly, several Grid programming and runtime environments differ considerably. However, even if some layer of homogenization could be enforced across various Grids, managing control programmatically through the various virtual organization will continue to linger [50].

## VI. THE POSSIBLE ECONOMIC IMPACT

It is our believe that on-demand grid provisioning through cloud infrastructure of grid test environments shall provide an effective result for attaining better Return On Investment (ROI) on assets. Cloud computing optimizes the number of concerns that plague recent methods to test environments. Cloud-based testing influences on-demand grid provisioning which can be more openly deployed, reducing the prerequisite principal expenses required to buy infrastructure (hardware and software). The pay-per-use paradigm offers a number of economic advantages, which comprises of providing testing organizations with autonomy from holding assets and reduced time-to-market for important commercial software. Most of these benefits are provided by leveraging access to cloud-based resources through the Cyberspace without necessarily compromising security requirements. Production-like environments at cheaper rates can be accomplished through on-demand grid provisioning using IaaS cloud, coupled with effective processes and governance that forms the discipline to cultivate superior images and version control. Establishments would perform better when they combine on-demand grid provisioning with business best practices like shift-left to develop the general quality of software and minimize the cost of application development, maintenance and testing [51].

Processing a data demanding process with different databases can produce an I/O logjam triggering the processor to execute uneconomically and thereby affecting commercial sustainability. This also relates to clouds computing but, because of the compacted nature of cloud environments and to the settings of cloud applications which recurrently move a moderately infrequent sizable data (with some exceptions), the issue is far less serious [52]

TABLE 1: SUMMARY TABLE OF TOOLS USED FOR ON-DEMAND GRID PROVISIONING USING CLOUD INFRASTRUCTURE

| No. | Reference | Grid Middleware | Virtualization Technology | Cloud Infrastructure | Grid-Cloud Integration Classification |
|---|---|---|---|---|---|
| 1 | Caron, et al. [2] | DIET-Solve | Xen | EUCALYPTUS | Disjointed/Full |
| 2 | De Assunção, et al. [53] | - | OpenNebula | EUCALYPTUS | Partial |
| 3 | Di Costanzo, et al. [14] | - | OpenNebula | Amazon EC2 | Partial |
| 4 | Dornemann, et al. [20] | Globus | - | Amazon EC2 | Partial |
| 5 | Jha, et al. [50] | - | Xen | Amazon EC2 & S3 | Partial |
| 6 | Nurmi, et al. [25] | - | Xen | EUCALYPTUS | - |
| 7 | Ostermann, et al. [35] | ASKALON | Xen | EUCALYPTUS | Partial |
| 8 | Smith, et al. [45] | Globus | Xen | - | - |
| 9 | Vázquez, et al. [44] | Globus | Xen & VMWare | Amazon EC2, GoGrid & Globus Nimbus | Partial |
| 10 | Caron, et al. [42] | - | - | Amazon EC2 | - |
| 11 | de Assunção, et al. [8] | - | OpenNebula | Amazon EC2 & EUCALYPTUS | Partial |
| 12 | Llorente, et al. [5] | Globus | OpenNebula | Amazon EC2 | Partial |
| 13 | Murphy, et al. [46] | Globus & Condor | KVM | VOC | Partial |
| 14 | Murphy and Goasguen [28] | Globus | KVM | VOC | Partial |
| 15 | Toorop and van Hoof [54] | Globus | - | Amazon EC2 | Partial |
| 16 | Wang, et al. [18] | GridSAM | Xen & VMWare | Amazon EC2 & Eucalyptus | Partial |
| 17 | Demchenko, et al. [37] | GEMBus | - | Composable Services Architecture (CSA) | Full |
| 18 | Mateescu, et al. [55] | Condor | OpenNebula | Amazon EC2 & Nimbus | Partial |
| 19 | Vázquez, et al. [3] | Globus | VMWare | Nimbus | Disjointed |
| 20 | Alvarez, et al. [38] | - | OpenNebula | CICA GRID solution | Full |
| 21 | Bosin [56] | - | OpenNebula | Eucalyptus | Partial |
| 22 | Chapman, et al. [10] | Condor | OpenNebula | RESERVOIR project | - |
| 23 | Rodero-Merino, et al. [16] | DIET-Solve | OpenNebula | Nimbus & Eucalyptus | Partial/Full |
| 24 | Saad, et al. [47] | Condor & BonjourGrid | Xen | Grid5000 | Partial |

| 25 | Nicolae and Rafique [57] | - | KVM | Grid5000 | - |
| 26 | Varma and Choi [12] | Globus | Xen & KVM | Eucalyptus | Partial |
| 27 | Casola, et al. [39] | Globus | Xen, KVM, OpenNebula & VMWare | Nimbus | Full |

**Key:** - (not specified)

## VII. DISCUSSION

From the Table 1 above, it shows that different Grid middlewares have been used to offer Grid resources to users in different researches demonstrating on-demand grid. But the most commonly used middleware is the Globus toolkit, which has been used in used in different experiments by the majority of the researchers [5, 20, 28, 46]. The Condor [10, 46, 55] is the second most relatively used, while the ASKALON [35], DIET-Solve [2], GridSAM [18], GEMBus [37] and BonjourGrid [47] have all been used at least ones to experiment different concepts and architectures in the on-demand Grid using Cloud infrastructures. This shows that Globus toolkit is the widely acceptable middleware and the most demanding by the users.

The Table 1 also shows that the two most common hypervisors used to host VMs in setting up the on-demand grid within the cloud are the Xen [2] and the OpenNebula [14]. These are closely followed by the KVM [28] and VMWare [44]. The combination of these virtualization technologies and the Cloud infrastructures technologies listed in the table above help to achieve the hosting and management of the VMs. The most frequently used Clouds for platforms for this purposes are Amazon EC2 [42] and the Eucalyptus [2]. Others are includes the GoGrid [44], VOC [28] and Nimbus [3]. It can also be seen that researches started in this area in the year 2009 and up till now more researches are being carried out in this emerging area of research.

We can also deduct from the Table 1above that, most of the researchers uses the Partial Grid-Cloud integration approach. This is in order to provide grid resources to users that already have a partial grid running their jobs and only needed additional grid resources at peak periods to take care of critical jobs. The some of the researchers uses the Full Grid-Cloud integration that takes care of a user whom does not have an existing grid and only need to run their jobs on-demand to save the cost of setting up a grid environment. The Full Grid-Cloud integration approach has the whole grid nodes set up with the infrastructural cloud and protected by it. Even though the Disjointed Grid-Cloud integration is not very popular, but some researchers have used it to verify the efficiency of their designs or to develop a new method.

## VIII. CONCLUSION AND FUTURE WORKS

Cloud computing technology presents the users with a number of virtualized resources for every task. The combination of Grid and Cloud computing presents even more power to organisations to provide Grid computing resources to the users on pay-per-use basis. In this paper, we survey various existing components used in the design of the on-demand grid resources provisioning using the Infrastructure as a service (IaaS) cloud. We presented an extended classification of virtualization technology [19] and a new taxonomy for the classification of Grid-Cloud integration. The table 1 above summarized the various tools used in previous researches to implements different architectures for the on-demand Grid resource provisioning in Cloud. They are presented here so that adequate care should be taken while designing future architectures for further researches.

On the basis of the above findings, the authors hereby recommend new or enhanced fault tolerance and scheduling algorithms to be designed for the on-demand grid environment. This is in order to take care of parameters like the economic cost, service level agreement (SLA) and interoperability issues.

## REFERENCES

[1] M. A. Vouk, "Cloud computing—Issues, research and implementations," in *Information Technology Interfaces, 2008. ITI 2008. 30th International Conference on*, 2008, pp. 31-40.

[2] E. Caron, F. Desprez, D. Loureiro, and A. Muresan, "Cloud computing resource management through a grid middleware: A case study with DIET and eucalyptus," in *Cloud Computing, 2009. CLOUD'09. IEEE International Conference on*, 2009, pp. 151-154.

[3] C. Vázquez, E. Huedo, R. S. Montero, and I. M. Llorente, "On the use of clouds for grid resource provisioning," *Future Generation Computer Systems*, vol. 27, pp. 600-605, // 2011.

[4] J. E. Smith and R. Nair, "The architecture of virtual machines," *Computer*, vol. 38, pp. 32-38, 2005.

[5] I. M. Llorente, R. Moreno-Vozmediano, and R. S. Montero, "Cloud computing for on-demand grid resource provisioning," *Advances in Parallel Computing*, vol. 18, pp. 177-191, 2010.

[6] N. Fallenbeck, H.-J. Picht, M. Smith, and B. Freisleben, "Xen and the art of cluster scheduling," in *Proceedings of the 2nd International Workshop on Virtualization Technology in Distributed Computing*, 2006, p. 4.

[7] R. Buyya and R. Ranjan, "Special section: Federated resource management in grid and cloud computing systems," *Future Generation Computer Systems,* vol. 26, pp. 1189-1191, 10// 2010.

[8] M. D. de Assunção, A. di Costanzo, and R. Buyya, "A cost-benefit analysis of using cloud computing to extend the capacity of clusters," *Cluster Computing,* vol. 13, pp. 335-347, 2010.

[9] M. Rosenblum and T. Garfinkel, "Virtual machine monitors: Current technology and future trends," *Computer,* vol. 38, pp. 39-47, 2005.

[10] C. Chapman, W. Emmerich, F. G. Márquez, S. Clayman, and A. Galis, "Software architecture definition for on-demand cloud provisioning," *Cluster Computing,* vol. 15, pp. 79-100, 2012.

[11] I. Foster, C. Kesselman, and S. Tuecke, "The anatomy of the grid: Enabling scalable virtual organizations," *International journal of high performance computing applications,* vol. 15, pp. 200-222, 2001.

[12] N. M. K. Varma and E. Choi, "Extending Grid Infrastructure Using Cloud Computing," in *Ubiquitous Information Technologies and Applications*, ed: Springer, 2013, pp. 507-516.

[13] I. Foster, Y. Zhao, I. Raicu, S. Y. Lu, and Ieee, *Cloud Computing and Grid Computing 360-Degree Compared*. New York: Ieee, 2008.

[14] A. Di Costanzo, M. D. De Assuncao, and R. Buyya, "Harnessing cloud technologies for a virtualized distributed computing infrastructure," *internet Computing, IEEE,* vol. 13, pp. 24-33, 2009.

[15] L. Wang, G. Von Laszewski, A. Younge, X. He, M. Kunze, J. Tao, *et al.*, "Cloud computing: a perspective study," *New Generation Computing,* vol. 28, pp. 137-146, 2010.

[16] L. Rodero-Merino, E. Caron, A. Muresan, and F. Desprez, "Using clouds to scale grid resources: An economic model," *Future Generation Computer Systems,* vol. 28, pp. 633-646, 4// 2012.

[17] M. D. ABDULMALIK and M. A. Shafi'i, "Windows Vista Kernel-Mode: Functions, Security Enhancements and Flaws," *Leonardo Journal of Sciences,* vol. 7, pp. 99-110, 2008.

[18] L. Wang, G. von Laszewski, M. Kunze, J. Tao, and J. Dayal, "Provide Virtual Distributed Environments for Grid computing on demand," *Advances in Engineering Software,* vol. 41, pp. 213-219, 2// 2010.

[19] J. Sahoo, S. Mohapatra, and R. Lath, "Virtualization: A survey on concepts, taxonomy and associated security issues," in *Computer and Network Technology (ICCNT), 2010 Second International Conference on*, 2010, pp. 222-226.

[20] T. Dornemann, E. Juhnke, and B. Freisleben, "On-demand resource provisioning for BPEL workflows using Amazon's elastic compute cloud," in *Cluster Computing and the Grid, 2009. CCGRID'09. 9th IEEE/ACM International Symposium on*, 2009, pp. 140-147.

[21] I. M. Abbadi, "Clouds' infrastructure taxonomy, properties, and management services," in *Advances in Computing and Communications*, ed: Springer, 2011, pp. 406-420.

[22] M. B. Belgacem, H. Hafsi, and N. Abdennadher, "A Hybrid Grid/Cloud Distributed Platform: A Case Study," in *Grid and Pervasive Computing*, ed: Springer, 2013, pp. 162-169.

[23] VMWare, "Understanding Full Virtualization, Paravirtualization, and Hardware Assist," *White Paper,* p. 4, 2007.

[24] J. Wu, R. Siewert, A. Hoheisel, J. Falkner, O. Strauß, D. Berberovic, *et al.*, "The Charité Grid Portal: User-friendly and Secure Access to Grid-based Resources and Services," *Journal of Grid Computing,* vol. 10, pp. 709-724, 2012/12/01 2012.

[25] D. Nurmi, R. Wolski, C. Grzegorczyk, G. Obertelli, S. Soman, L. Youseff, *et al.*, "The eucalyptus open-source cloud-computing system," in *Cluster Computing and the Grid, 2009. CCGRID'09. 9th IEEE/ACM International Symposium on*, 2009, pp. 124-131.

[26] G. B. Barone, R. Bifulco, V. Boccia, D. Bottalico, R. Canonico, and L. Carracciuolo, "GaaS: Customized grids in the clouds," vol. 7640 LNCS, ed, 2013, pp. 577-586.

[27] L. Nie and Z. W. Xu, "An Adaptive Scheduling Mechanism for Elastic Grid Computing," *2009 Fifth International Conference on Semantics, Knowledge and Grid (Skg 2009),* pp. 184-191, 2009.

[28] M. A. Murphy and S. Goasguen, "Virtual Organization Clusters: Self-provisioned clouds on the grid," *Future Generation Computer Systems,* vol. 26, pp. 1271-1281, 2010.

[29] P. A. D. Figueiredo, J.A.B. Fortes, "A case for grid computing on virtual machines," *23rd International Conference on Distributed Computing Systems,* 2003.

[30] S. Manavi, S. Mohammdalian, N. I. Udzir, and A. Abdullah, "Secure Model for Virtualization Layer in Cloud Infrastructure," *International Journal of Cyber-Security and Digital Forensics (IJCSDF),* vol. 1, pp. 32-40, 2012.

[31] C.-T. Yang, K.-L. Huang, W. C.-C. Chu, F.-Y. Leu, and S.-F. Wang, "Implementation of Cloud IaaS for Virtualization with Live Migration," in *Grid and Pervasive Computing*, ed: Springer, 2013, pp. 199-207.

[32] V. Sharma, "Virtual machine management in cloud computing environment," 2013.

[33] G. Aceto, A. Botta, W. de Donato, and A. Pescapè, "Cloud monitoring: A survey," *Computer Networks,* // 2013.

[34] D. Karakasilis, F. Georgatos, T. Alexopoulos, and L. Lambrinos, "Application of live video streaming over GRID and cloud infrastructures," 2011, pp. 379-383.

[35] S. Ostermann, R. Prodan, and T. Fahringer, "Extending grids with cloud resource management for scientific computing," in *Grid Computing, 2009 10th IEEE/ACM International Conference on*, 2009, pp. 42-49.

[36] C. G. Chaves, D. M. Batista, and N. L. S. da Fonseca, "Scheduling Grid Applications on Clouds," *2010 Ieee Global Telecommunications Conference Globecom 2010,* 2010.

[37] Y. Demchenko, C. De Laat, J. Van Der Ham, M. Ghijsen, V. Yakovenko, and M. Cristea, "On-demand provisioning of Cloud and Grid based infrastructure services for collaborative projects and groups," 2011, pp. 134-142.

[38] M. Alvarez, A. Fernandez-Montes, J. Ortega, and L. Gonzalez-Abril, "THE CICA GRID: A Cloud Computing Infraestructure on demand with Open Source Technologies," 2012.

[39] V. Casola, A. Cuomo, M. Rak, and U. Villano, "The CloudGrid approach: Security analysis and performance evaluation," *Future Generation Computer Systems,* vol. 29, pp. 387-401, // 2013.

[40] F. Leymann, "Cloud Computing," *it-Information Technology,* vol. 53, pp. 163-164, 2011.

[41] R. Buyya, C. S. Yeo, and S. Venugopal, "Market-oriented cloud computing: Vision, hype, and reality for delivering it services as computing utilities," in *High Performance Computing and Communications, 2008. HPCC'08. 10th IEEE International Conference on*, 2008, pp. 5-13.

[42] E. Caron, F. Desprez, and A. Muresan, "Forecasting for grid and cloud computing on-demand resources based on pattern matching," in *Cloud Computing Technology and*

*Science (CloudCom), 2010 IEEE Second International Conference on,* 2010, pp. 456-463.

[43] F. Hu, M. Qiu, J. Li, T. Grant, D. Taylor, S. McCaleb, *et al.*, "A Review on cloud computing: Design challenges in Architecture and Security," *Journal of Computing and Information Technology,* vol. 19, pp. 25–55, 2011.

[44] C. Vázquez, E. Huedo, R. S. Montero, and I. M. Llorente, "Dynamic provision of computing resources from grid infrastructures and cloud providers," in *Grid and Pervasive Computing Conference, 2009. GPC'09. Workshops at the,* 2009, pp. 113-120.

[45] M. Smith, M. Schmidt, N. Fallenbeck, T. Dörnemann, C. Schridde, and B. Freisleben, "Secure on-demand grid computing," *Future Generation Computer Systems,* vol. 25, pp. 315-325, 3// 2009.

[46] M. A. Murphy, L. Abraham, M. Fenn, and S. Goasguen, "Autonomic clouds on the grid," *Journal of Grid Computing,* vol. 8, pp. 1-18, 2010.

[47] W. Saad, H. Abbes, C. Cerin, and M. Jemni, "A Self-Configurable Desktop Grid System On-Demand," in *P2P, Parallel, Grid, Cloud and Internet Computing (3PGCIC), 2012 Seventh International Conference on,* 2012, pp. 196-203.

[48] M. Koehler, M. Ruckenbauer, I. Janciak, S. Benkner, H. Lischka, and W. N. Gansterer, "A grid services cloud for molecular modelling workflows," *International Journal of Web and Grid Services,* vol. 6, pp. 176-195, 2010.

[49] R. B. Asadzadeh et al, Chun Ling Kei, Deepa Nayar, and Srikumar Venugopal, "Global Grids and Software Toolkits: A Study of Four Grid Middleware

Technologies," *High-Performance Computing: Paradigm and Infrastructure,* 2005.

[50] S. Jha, A. Merzky, and G. Fox, "Using clouds to provide grids with higher levels of abstraction and explicit support for usage modes," *Concurrency and Computation: Practice and Experience,* vol. 21, pp. 1087-1108, 2009.

[51] V. K. Mylavarapu, "Why On-Demand Provisioning Enables Tighter Alignment of Test and Production Environments," *Cognizant Reports,* pp. 1-6, 2011.

[52] M. Cafaro and G. Aloisio, *Grids, Clouds, and Virtualization:* Springer, 2011.

[53] M. D. De Assunção, A. Di Costanzo, and R. Buyya, "Evaluating the cost-benefit of using cloud computing to extend the capacity of clusters," in *Proceedings of the 18th ACM international symposium on High performance distributed computing,* 2009, pp. 141-150.

[54] W. Toorop and A. van Hoof, "On Demand Grid on Cloud," 2010.

[55] G. Mateescu, W. Gentzsch, and C. J. Ribbens, "Hybrid computing—where HPC meets grid and cloud computing," *Future Generation Computer Systems,* vol. 27, pp. 440-453, 2011.

[56] A. Bosin, "An SOA-based model for the integrated provisioning of cloud and grid resources," *Adv. Soft. Eng.,* vol. 2012, pp. 12-12, 2012.

[57] B. Nicolae and M. Rafique, "Leveraging Collaborative Content Exchange for On-Demand VM Multi-Deployments in IaaS Clouds," in *Euro-Par'13: 19th International Euro-Par Conference on Parallel Processing,* 2013.